\documentclass[conference]{IEEEtran}
\IEEEoverridecommandlockouts
\usepackage{cite}
\usepackage{amsmath,amssymb,amsfonts}
\usepackage{graphicx}
\usepackage{textcomp}
\usepackage{xcolor}
\usepackage{comment}
\usepackage[T1]{fontenc}
\usepackage[utf8]{inputenc}
\usepackage{dblfloatfix} % or stfloats
\usepackage{algorithm}
\usepackage{algpseudocode}
\usepackage{subcaption}
\usepackage{soul}
\soulregister\mathcal{1}
\soulregister\subseteq{1}
\usepackage{xcolor}
\usepackage{graphicx}

\def\BibTeX{{\rm B\kern-.05em{\sc i\kern-.025em b}\kern-.08em
    T\kern-.1667em\lower.7ex\hbox{E}\kern-.125emX}}
\begin{document}

\title{Load Balancing in Non-Terrestrial Networks Using Free Space Optical Inter-satellite Links
}
\author{%
\IEEEauthorblockN{%
Abid Afridi\textsuperscript{*},
Alexis A. Dowhuszko\textsuperscript{*},
Jevgenij Krivochiza \textsuperscript{\dag},
Risto Wichman\textsuperscript{*},
Jyri H\"am\"al\"ainen\textsuperscript{*}}
\IEEEauthorblockA{\textsuperscript{*}Department of Information and Communications Engineering, Aalto University, Finland.}
\IEEEauthorblockA{\textsuperscript{\dag}Department of Product and Innovation, SES, 6815 Betzdorf, Luxembourg} 
\IEEEauthorblockA{Emails: \{abid.afridi, alexis.dowhuszko, risto.wichman, jyri.hamalainen\}@aalto.fi; \{jevgenij.krivochiza\}@ses.com} \vspace{-7mm}
% Also this one in the final �� �� 
% Probably EU wants an acknowledgement as well
\thanks{This work was partially funded by the FOCAL, a doctoral network of the EU’s Horizon Europe programme in the frame of the MSCA GA n°101169042 and partially supported by Research Council of Finland Grant 339446. Views and opinions expressed do not necessarily reflect those of the SES.}
}
\maketitle
\begin{abstract}
Non-terrestrial networks (NTNs) increasingly rely on non-geostationary (NGSO) constellations that combine radio frequency (RF) feeder links (FLs) with free space optical (FSO) inter-satellite links (ISLs). Downlink performance in such systems is often constrained by uneven satellite-gateway visibility, data traffic congestion, and rain-induced FL attenuation, leaving the downlink capacity of some satellites underutilized while others become bottlenecks. To prevent such non-uniform load distribution, this paper presents a fairness-driven load balancing strategy that treats the satellite constellation in space as an anycast multi-commodity flow problem. Then, by solving an equivalent linear programming optimization problem, the proposed algorithm dynamically selects the most convenient ground station (GS) to serve each satellite and, when needed, offloads data traffic to adjacent satellites through FSO ISLs. Using a realistic MEO satellite constellation with 1550 nm FSO ISLs and Ka-band feeder links, the method stabilizes the reverse link data service, maintaining the average data rate but notably improving the worst-case throughput. Our proposed algorithm enhances the minimum downlink data rate by more than 25\% in the presence of rain and by over 10\% under no-rain conditions. These results demonstrate that the use of an ISL-assisted load-balancing scheme mitigates FL bottlenecks and enhances fairness across the satellite constellation, offering a scalable basis for resource allocation in future NTN systems.
\end{abstract}
\begin{IEEEkeywords}
Non-terrestrial network, medium earth orbit, anycast multi-commodity flow problem, optical inter-satellite link, linear programming. 
\end{IEEEkeywords}
\vspace{-1mm}
\section{Introduction}
Non-terrestrial networks (NTNs) play a pivotal role in beyond-5G/6G systems, delivering global connectivity and complementing terrestrial infrastructure, which includes unmanned aerial vehicles (UAVs), high-altitude platforms (HAPs), and mega-constellations of satellite systems. NTNs have been historically deployed for targeted missions such as disaster response, navigation, television broadcasting, and remote sensing. However, recent tremendous developments of aerial/space technologies, coupled with reduced costs of their manufacturing and launching, have
enabled more applications of NTNs when integrated with terrestrial communication networks \cite{ortiz2025gen}. 

Among NTNs, the choice of the satellite orbits reflect tradeoffs in latency, coverage, mobility, and control. Geostationary earth orbit~(GEO) satellites (35786 km) provide wide and stable coverage on a global scale with no handovers, but suffer from high path loss and notable end-to-end delays in the order of ~270 ms, limiting interactive services. At the other extreme, low earth orbit (LEO) satellites (500-2000~km) minimize the propagation latency from space to earth,  but their relatively small footprints and fast motion demand large constellations, frequent handovers, and complex inter-satellite routing \cite{krivochiza2021end,lyras2019optimizing}. In the middle of both cases, medium earth orbit (MEO) satellites (8000–20000 km) offer a trade-off solution, reducing delay relative to GEO while offering broader service areas and slower motion than LEO, which lowers handover frequency and simplifies feeder-link (FL) geometry and backhaul planning. With fewer satellites, MEO satellites support high-rate space links and scalable multi-orbit architectures. Other 3 billion (O3b) systems, for instance, have demonstrated make-before-break handovers and stable Ka-band backhaul for 5G nodes, underscoring the role of MEO as a practical middle ground for integrated NTNs \cite{chougrani2024connecting}.

As traffic demand, weather conditions, and satellite-GS visibility vary over time, per-satellite throughput can become highly uneven, with some satellite nodes starved by local FL constraints while others enjoy headroom \cite{ning2023load}. Under bad weather conditions, FLs degrade significantly, and  attenuation increases with rain rate most strongly at higher frequencies, such as the Ka band \cite{shrestha2016study}. Considering this need, next-generation MEO constellations increasingly employ inter-satellite links (ISL) as a transport layer to redistribute load in space rather than forwarding all traffic directly to ground stations (GS) \cite{10924685}. Optical ISLs are especially attractive, as they offer higher data rates, much smaller apertures (lower mass/volume), narrower beams that suppress interference and improve security, and reduced transmit power owing to low beam spread and high directivity \cite{chaudhry2020free} \cite{dowhuszko2019total}. Related studies on LEO/MEO ISLs report multi-Gb/s capabilities and explore modulation/format choices that make them attractive for backhaul within the constellation~\cite{chaudhry2021laser}.
\begin{figure*}[h]
  \centering
  \begin{subfigure}[t]{0.56\textwidth}
    \centering
    \includegraphics[width=\linewidth]{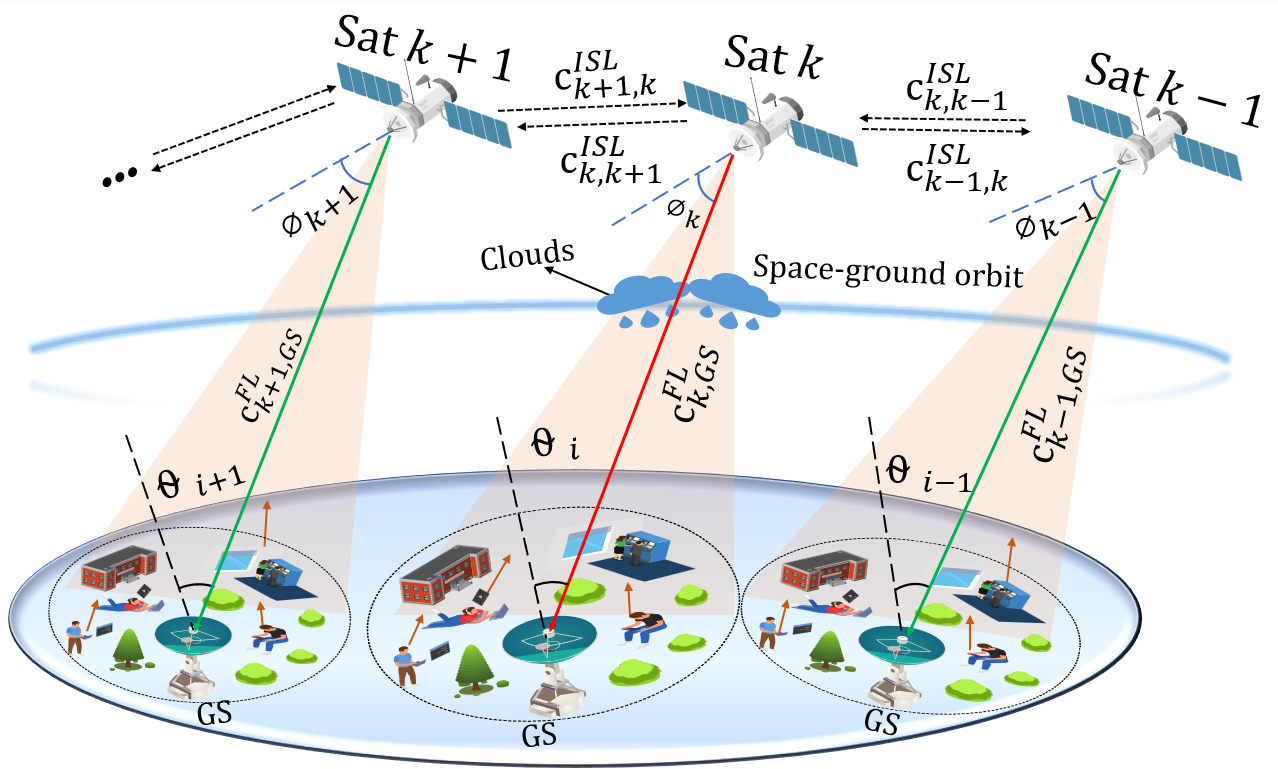}
    \caption{}
        \label{fig:system}
  \end{subfigure}
  \hfill
  \begin{subfigure}[t]{0.42\textwidth}
    \centering
    \includegraphics[width=\linewidth]{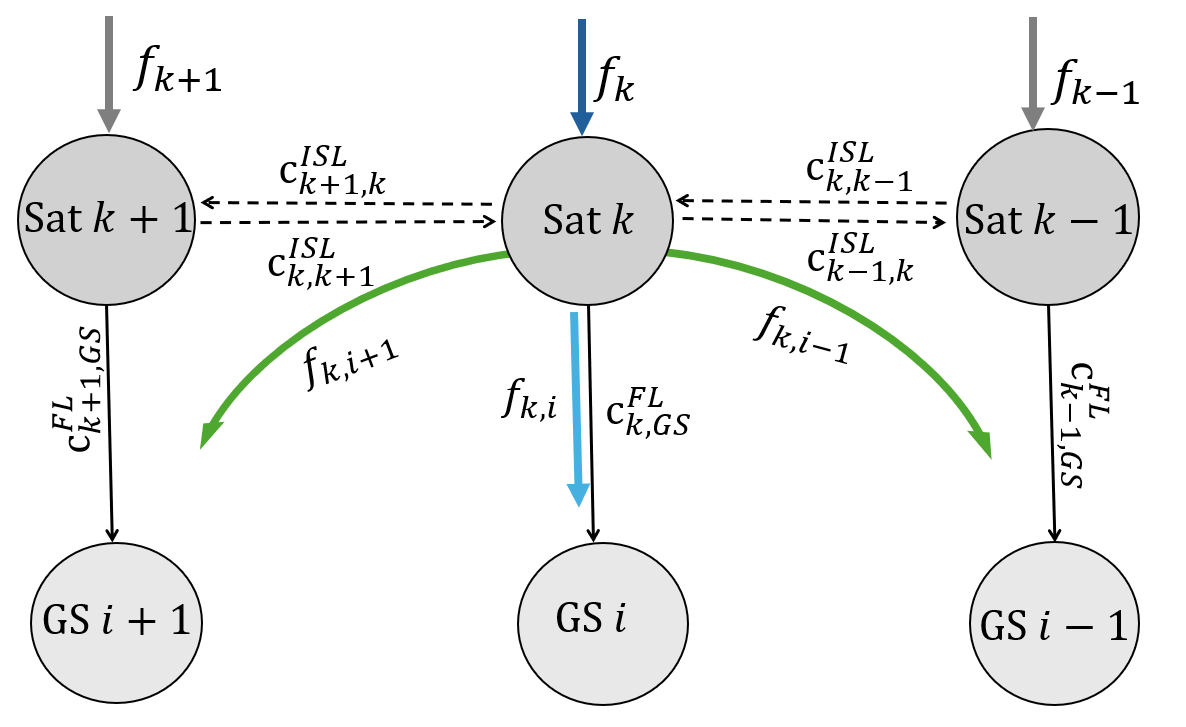}
    \caption{}
        \label{fig:comm}
  \end{subfigure}
  \caption{\footnotesize \textbf{(a)} Overview of a multi-satellite constellation with \(K\) satellites serving \(I\) ground stations. Each satellite forms a single Ka-band feeder link to its serving GS. The achievable FL capacity varies with geometry, off-axis angles (\(\phi\) and \(\theta\)), and rain attenuation. The red FL indicates a weather-degraded connection affected by rain, whereas the green FLs represent no-rain conditions. The adjacent satellites are connected by optical ISL with fixed capacity. These ISLs offload traffic from satellites with weak feeder links to neighbors with stronger feeder links, improving constellation download rate and fairness. \textbf{(b)} Equivalent representation of the multi-satellite system in the form of a time-varying graph. Satellite \(k\) downloads data to GS \(i\) directly or via neighboring satellites (with satellite indexes \(k\pm1\)) using ISLs.}
  \label{fig:system-comm}\vspace{-3mm}
\end{figure*}

Most of the work reported in the literature so far has focused on developing optimal resource allocation algorithms to schedule data exchange from satellites to a single or multiple GSs, such that the total throughput of data downloading is maximized \cite{spangelo2015optimization}. In-space constellation routing shows that the concentration of GS produces moving bottlenecks in the space segment near feeder nodes, degrading utilization when shortest-path routing is used without explicit load balancing. Classic path-based designs for LEO/MEO with ISLs (e.g., time-sliced shortest paths or precomputed virtual topologies) handle these dynamics but typically optimize for reachability or total throughput, not fairness across spacecraft under shared FL/ISL budgets \cite{werner1997atm}. Finally, some papers considered load distribution assuming a single pre-defined GS sink for all the satellites under consideration\cite{jia2017collaborative}.

\setlength{\parskip}{0pt}
In this paper, we study the benefits of adding optical ISLs to a MEO satellite constellation that supports traffic offloading under variability in the capacity of individual feeder links. The system-level performance gain of a MEO satellite constellation is evaluated using the max-min rate per satellite objective function in different weather conditions. To solve this optimization problem, the satellite constellation is modeled as a time-varying graph, and an anycast multi-commodity flow problem (ACMCFP) is formulated in which the data traffic of each satellite can be delivered to any GS either directly over its FL or via one-hop offloading simultaneously to a neighboring satellite over an ISL. The optimization problem is computed using linear programming. Obtained simulation results show that ISL-assisted load balancing improves fairness by increasing the minimum per-satellite downloaded throughput relative to a comparable MEO constellation without ISLs, particularly during rain conditions.

The rest of the paper is organized as follows: Section II presents the system model for the MEO satellite constellation, including the formulas that estimate the download data rate per satellite. Section~III describes the problem
formulation and proposed solution for ACMCFP, which includes the algorithm to solve the max-min problem. The parameters of the simulation setting, as well as the performance analysis of the obtained results, are presented in Section IV. Finally, conclusions are discussed in Section V.
\section{System Model}
We consider a satellite communication network consisting of a constellation of \textit{K} satellites with the indexes ${k}$ in set $\mathcal{K}$ and \textit{I} GSs with the indexes ${i}$ in set $\mathcal{I}$. The communication architecture supports two types of links: FL connecting satellites to terrestrial infrastructure and optical ISL enabling direct communication between satellites, as shown in Fig.~\ref{fig:system}. The network operates over a discrete time horizon divided into \textit{N} slots, with time indexes in the set $\mathcal{T} = \{1, 2, \ldots, \textit{N}\}$. Each time slot represents a period during which the satellite network topology remains quasi-static, allowing for deterministic resource allocation decisions. The data flow originating from satellite \(k\) towards any reachable GS \(i\) can be expressed as
\begin{align}
    f_k &= f_{k,i} + \sum_{j \in \mathcal{I}} f_{k,j},
\end{align} 
where \(f_{k,i}\) denotes the direct flow from satellite $k$ to GS $i$, and the summation on the right-hand side accounts for possible relay flows toward other neighboring GSs. We assume that adjacent satellites are always connected via ISLs with fixed link capacities, as shown in Fig. \ref{fig:comm}. The fraction of capacity allocated to a particular FL or ISL is denoted by \(w\) and \(v\), respectively.

\subsection*{A. Feeder Link Model for Downlink}
The received power at GS \(i\) from the satellite \(k\) in the time slot \(n\) is given by  
\begin{equation}
P_{i}[n] \;=\; \frac{P_k \, g_{k}\!\left(\phi_{k,i}[n]\right) \, g_{i}\!\left(\theta_{i,k}[n]\right)}{L_{T}[n]},
\label{eq:Pr_phi}
\end{equation}
where \(P_k\) is the transmit power of satellite \(k\), \(g_{k}\!\left(\phi_{k,i}[n]\right)\) denotes the transmit antenna gain of the satellite \(k\) in the direction of the GS \(i\) with off-axis angle \(\phi_{k,i}[n]\), and \(g_{i}\!\left(\theta_{i,k}[n]\right)\) represents the receiving antenna gain, parameterized by the receiving angle \(\theta_{i,k}[n]\). For simplicity, we assume that the antenna in the GS has the possibility to follow the trajectory of the satellite, enabling an almost perfect alignment such that \(\theta_{i,k}[n] = 0\) can be assumed. The total path loss attenuation between satellite~$k$ and GS \(i\) at time \(n\) can be then written as  
\begin{equation}
L_{T}[n] = \bar{L}_{k,i}[n]\; L_{\text{sf}}\;L_{\text{rain}},
\label{eq:Ltotal}
\end{equation}
where \(\bar{L}_{k,i}[n] = \left(\frac{4\pi d_{k,i}}{\lambda}\right)^2\) denotes the free space path loss~(FSPL), with \(d_{k,i}[n]\) representing the distance between satellite~$k$ and GS \(i\), and \(\lambda \) is the radio frequency carrier wavelength \cite{ITU_P525}. The term \(L_{\text{sf}}\) in (3) accounts for large-scale shadowing. The attenuation that rain absorption introduces in case of bad weather conditions can be expressed as
\begin{align}
    L_\text{rain} = \left(d_{\text{eff}} \cdot \gamma_{\text{rain}}\right),
\end{align}
where $d_{\text{eff}}$ represents the effective distance, which depends on the actual 
path length and the long-term statistical rain rate $\rho_{0.01}$, 
which represents the rain rate exceeded for $0.01\%$ of the time \cite{11176007}. To determine this value, \(\gamma_{\text{rain}} \; = a \cdot \rho^b\) shows specific attenuation, with  $\rho$ representing the rain rate, and $a$ and $b$ are empirical constants influenced by frequency, polarization, and elevation angle. For 20~GHz and horizontal polarization, ITU-R P.838-3 recommends to use $a = 0.09164$ and $b = 1.0568$~\cite{ITU-RP.838-3}. %These three sample cities were selected because the O3b satellite system has GSs there, as shown in Fig. \ref{fig:ses-system}. 
\begin{figure}[t]   % or [htbp]
  \centering
  \includegraphics[width=\columnwidth]{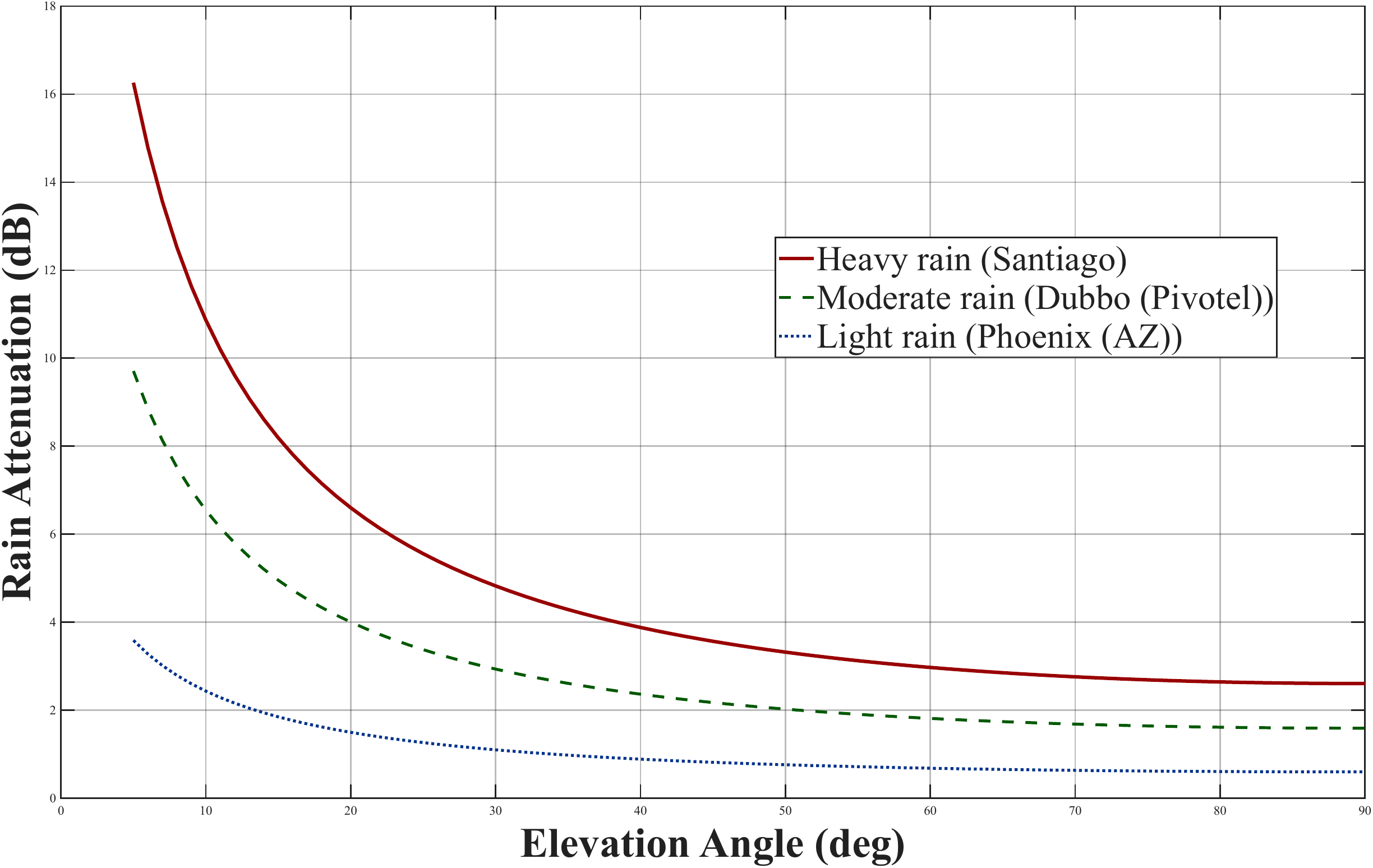}
  \caption{\footnotesize Rain attenuation (dB) across different elevation angles.}
  \label{fig:Boxplots}\vspace{-4mm}
\end{figure}

 As shown in Fig. \ref{fig:Boxplots}, the rain attenuation varies as a function of elevation angle for three the rainfall conditions under study in this paper: heavy rain (Santiago), moderate rain (Dubbo), and light rain (Phoenix). At low elevation angles below 10°, the attenuation is most severe, reaching values above 16 dB under heavy rain conditions. As the elevation angle increases, the slant-path length through the rain medium decreases, resulting in a monotonic decrease in attenuation. For heavy rain, attenuation gradually drops to about 3 dB at 80° elevation, while moderate and light rain scenarios show proportionally lower losses, converging to values near 2 dB and 1 dB, respectively. These results highlight the strong dependency of rain-induced attenuation on elevation angle, with lower angles posing the greatest challenge for link reliability.
 The three sample cities were selected because the O3b satellite system has GSs in each of them, as shown in Fig. \ref{fig:ses-system}. 

\subsection*{B. Optical Inter-Satellite Link Model}
An optical ISL is the link between two satellites, and the
propagation medium is the vacuum of space. For an optical ISL, the
received power from the satellite $k$ at the adjacent satellite $\ell$ at a time slot \(n\) is given by:
\begin{equation}
    P_\ell [n] = P_k[n] \eta_k \eta_\ell G_k[n] G_\ell[n] L_k[n] L_\ell[n] L_{\text{ps}[n], 
    \label{eq:PR}}
\end{equation}
where $P_k$ is the transmitted
power in watts, $(\eta_k,\eta_\ell)$ is the optics efficiency of the transmitter and receiver, $(G_k,G_\ell)$ is the
transmitter and receiver gain, $(L_k,L_\ell)$ is the transmitter and receiver pointing loss, and $L_{\text{ps}}=\left(\frac{\lambda}{4\pi d_{\rm isl}}\right)^2$ represents the FSPL for the optical link between satellites where $\lambda$ is the operating wavelength in nm, and $d_{\rm isl}$ is the
distance between satellites in km \cite{liang2022link}. The
transmitter gain is given by
$G_k = 16/(\Theta_k)^2$, where $\Theta_k$ is the full transmitting
divergence angle in radians. Similarly, the receiver gain is expressed
as $G_\ell = (D_\ell \pi/\lambda)^2$, where $D_\ell$ is the receiver's telescope
diameter in mm; the transmitter pointing loss is given as
$L_k=\exp\!\big(-G_k(\Phi_k)^2\big)$, where $\Phi_k$ is the
transmitter pointing error in radians; the receiver pointing loss
is written as $L_\ell=\exp\!\big(-G_\ell(\Phi_\ell)^2\big)$, where
$\Phi_\ell$ is the receiver pointing error in radians.

\subsection*{C. Aggregate Data Rate per Satellite with ISL}
The instantaneous carrier-to-noise ratio (CNR) for the FL from satellite $k$ to GS $i$ can be expressed as
\begin{equation}
\gamma^{\text{FL}}_{k,i}[n]
= \frac{P_{i}[n]}{N_i[n]},
\label{eq:gamma_fl}
\end{equation}
where \(N_i[n]=k_B\, T_{i}\, B_{i}[n]\) defines the noise power with \(k_B\) denoting Boltzmann’s constant, \(T_i\) the system noise temperature (in Kelvin), and \(B_{i}[n]\) the noise-equivalent bandwidth~\cite{3GPP_TR38821}.  While, for ISL between satellite \(k\) and \(\ell\), the received CNR can be calculated as
\begin{equation}
\gamma^{\text{ISL}}_{k,\ell}[n]
= \frac{P_{\ell}[n]}{\sigma_\ell[n]},
\label{eq:gamma_isl}
\end{equation}
where \(P_{\ell}[n]\) is the received power signal at satellite \(\ell\) and \(\sigma_\ell[n]\) is the noise power. Hence, the achievable data rates for FL and optical ISL can be approximated by the upper bound given by the Shannon's Formula, i.e.,
\begin{equation}
c_{k,i/\ell}^{\text{FL/ISL}}[n] = BW^{\text{FL/ISL}}_{k,i/\ell}[n]\log_2\!\left(1 +{\gamma^{\text{FL/ISL}}_{k,i/\ell}}[n]\right).
\label{eq:shannon_capacity}
\end{equation}
Hence, the aggregate download rate for satellite \(k\) at time \(n\) can be written as
\begin{multline}
R_k[n]=\underbrace{w_{k,i}^{(k,i)}[n] \cdot c_{k,i}^{\mathrm{FL}}[n]}_{{r_{k,i}^{(k,i)}}}\\
 + \sum_{\ell\in\mathcal{L}_k}\sum_{\substack{j\in \mathcal{J}_k \\ i\neq j}}
\text{min}  
\Big[ \underbrace{ v_{k,\ell}^{{(k,j)}}
[n] \cdot c_{k,\ell}^{\mathrm{ISL}}[n] {,}\;
w_{\ell,j}^{(k,j)}[n] \cdot c_{\ell,j}^{\mathrm{FL}}[n]}_{r_{k,j}^{(k,j)}} \Big] \quad \forall n,
\end{multline}
where the sub-set \(\mathcal{L}_k \subseteq \mathcal{K}\) contains the indexes of neighboring satellites directly connected to satellite \(k\) via ISLs, and the sub-set \(\mathcal{J}_k \subseteq \mathcal{I}\) contains the indexes of the GSs that can be reached by satellite \(k\) with one inter-satellite hop.  Thus, the fraction of the FL capacity allocated from satellite \(k\) to GS \(i\) is given by
\begin{equation}
        w_{k,i}=\sum_{\substack{S \in \mathcal{K}}}w_{k,i}^{(S,i)},
\end{equation}
where the superscripts \((S,i)\) denote the source and destination of the flow of data that is being transmitted, while the subscripts $(k, i)$ identify the physical transmitter–receiver link. Hence, the sum throughput of flows over FL from satellite~$k$ to GS \(i\) and over ISL from satellite \(k\) to satellite \(\ell\) is then obtained as
\begin{align}
r_{k,i}^{\text{FL}}[n]   &= w_{k,i}[n] \cdot c_{k,i}^{\text{FL}}[n], 
&& \forall k \in \mathcal{K},\; i \in \mathcal{I},\; n \in \mathcal{T}, 
\label{eq:capacity_utilization} \\[4pt]
r_{k,\ell}^{\text{ISL}}[n] &= v_{k,\ell}[n] \cdot c_{k,\ell}^{\text{ISL}}[n], 
&& \forall k,\ell \in \mathcal{K},\; n \in \mathcal{T},
\label{eq:ISL_capacity_utilization}
\end{align}
where \(w_{k,i}[n]\) and \(v_{k,\ell}[n]\) \(\in[0,1]\) denote the fraction of the FL and ISL capacities allocated at time slot \(n\), respectively.

\section{Problem Formulation and Proposed Solution}
The primary objective is to ensure fair distribution of data across all satellites by adopting a max–min fairness criterion. Specifically, the aim is to maximize the minimum download rate achieved by any satellite in the constellation at each time instant. To this end, we formulate the following optimization problem: \vspace{-2mm}
\begin{multline}
\max_{\{w_{k,i}^{(S,D)}[n]\},\,\{v_{k,\ell}[n]\}}
\min_{k\in\mathcal{K}}
\bigg(
R_k[n]
\bigg)
\ \forall n,
\label{eq:objective_global}
\end{multline}
verifying simultaneously the following constraints:
\begin{subequations}
\renewcommand{\theequation}{\theparentequation\alph{equation}}
\begin{multline}
R_{k}[n]- r_{k,i}^{\mathrm{FL}}[n]
\;-\; \\
\sum_{\ell\in\mathcal L_k} r_{k,\ell}^{\mathrm{ISL}}[n]
\;+\; \sum_{\ell\in\mathcal L_k} r_{\ell,k}^{\mathrm{ISL}}[n]
\;=\; 0, 
\quad \forall k\in\mathcal K,\; n\in\mathcal T, 
\tag{13a}
\end{multline}
\begin{equation}
 w_{k,i}[n] \;\geq\; 0,
\quad \forall k \in \mathcal{K},\; i \in \mathcal{I},\; n \in \mathcal{T},
\tag{13b}
\end{equation}
% --- ISL allocation bounds ---
\begin{equation}
0 \;\le\; v_{k,\ell}[n] \;\le\; 1,
\quad \forall k,\ell \in \mathcal{K},\; n \in \mathcal{T},
\tag{13c}
\end{equation}
% --- Per-link FL capacity fraction ---
\begin{equation}
w_{k,i}^{(k,i)}[n] + \sum_{\substack{\ell \in \mathcal{L}_k }} w_{k,i}^{(\ell,i)}[n]  \;\le\; 1,
\quad \forall k \in \mathcal{K},\; n \in \mathcal{T},
\tag{13d}
\end{equation}
\end{subequations}
In this formulation, \(c_{k,i}^{\text{FL}}[n]\) and \(c_{k,\ell}^{\text{ISL}}[n]\) denote the achievable 
data rates of FL and optical ISL, respectively, while 
\(r_{k,i}^{\text{FL}}[n]\) and \(r_{k,\ell}^{\text{ISL}}[n]\) represent the corresponding effective data 
rates after resource allocation. 
Constraint (13a) enforces flow conservation at every relay satellite, ensuring that the total outgoing traffic (towards GSs and neighboring satellites) becomes equal to the incoming traffic from neighboring satellites. Constraint (13b) ensures non-negativity of the FL allocation fractions. Constraint (13c) bounds the ISL allocation fractions within the interval \([0,1]\). Constraint (13d) guarantees that the aggregate fraction of FL capacity allocated to multiple flows sharing the same physical FL does not exceed unity in any time slot.
\begin{algorithm}[t]
\caption{Load balancing using the dual-simplex algorithm}
\label{alg:acmcfp}
\begin{algorithmic}[1]
\State \textbf{Input:}
Satellite set $\mathcal{K}$, Ground stations (GS) set $\mathcal{I}$, time slots $n\in\mathcal{T}$,
neighboring satellites set $\{\mathcal{L}_k\}_{k\in\mathcal{K}}$.
\State \textbf{Output:}
Per-satellite throughput $R_k[n]$,
flows $r^{\text{FL}}_{k,i}[n]$, $r^{\text{ISL}}_{k,\ell}[n]$,
scheduling weights $w_{k,i}[n]$, $v_{k,\ell}[n]$.
\For{each time slot $n \in \mathcal{T}$}
    \State Compute distances $d_{k,i}[n]$ and $d_{k,\ell}^{\text{ISL}}[n]$ based on GS and satellite positions.
    \State Compute off-axis angles $(\phi$ and $\theta)$ using (2).
    \State Compute achievable data rates for FL $c^{\text{FL}}_{k,i}[n]$ and ISL $c^{\text{ISL}}_{k,\ell}[n]$ using \eqref{eq:shannon_capacity}.
    \State Solve (13) using dual-simplex \texttt{linprog} to obtain optimal $w_{k,i}^\star[n]$, $v_{k,\ell}^\star[n]$.
\EndFor
\State \Return $w_{k,i}^\star[n]$, $v_{k,\ell}^\star[n]$, $r^{\text{FL}\star}_{k,i}[n]$, $r^{\text{ISL}\star}_{k,\ell}[n]$, $R_k^\star[n]$. 
\end{algorithmic} 
\end{algorithm}

The scheduling algorithm runs at each time slot \(n \in \mathcal{T}\) and computes the distances $d_{k,i}[n]$, $d_{k,\ell}^{\text{ISL}}[n]$ based on GS satellite positions. Then, the algorithm computes the line of sight visibilities, which includes the off-axis angles $(\phi$ and $\theta)$ using (2). After determining the off-axis angles, the algorithm proceeds to compute achievable 
data rates for FL $c^{\text{FL}}_{k,i}[n]$ and ISL $c^{\text{ISL}}_{k,\ell}[n]$ using link-budget analysis presented in \eqref{eq:shannon_capacity}. The algorithm then solves the optimization problem in (13)  with MATLAB \texttt{linprog} which uses the dual simplex algorithm to obtain the optimal fractions $w_{k,i}^\star[n], v_{k,\ell}^\star[n]$. This process continues for each time slot \(n \in \mathcal{T}\) to obtain the optimal solution in terms of $w_{k,i}^\star[n], v_{k,\ell}^\star[n], r^{\text{FL}\star}_{k,i}[n], r^{\text{ISL}\star}_{k,\ell}[n], R_k^\star[n]$. The pseudocode for the load balancing using the dual-simplex algorithm is summarized in Algorithm 1.

\section{Simulation results}
\begin{figure}[t]   % or [htbp]
  \centering
  \includegraphics[width=\columnwidth]{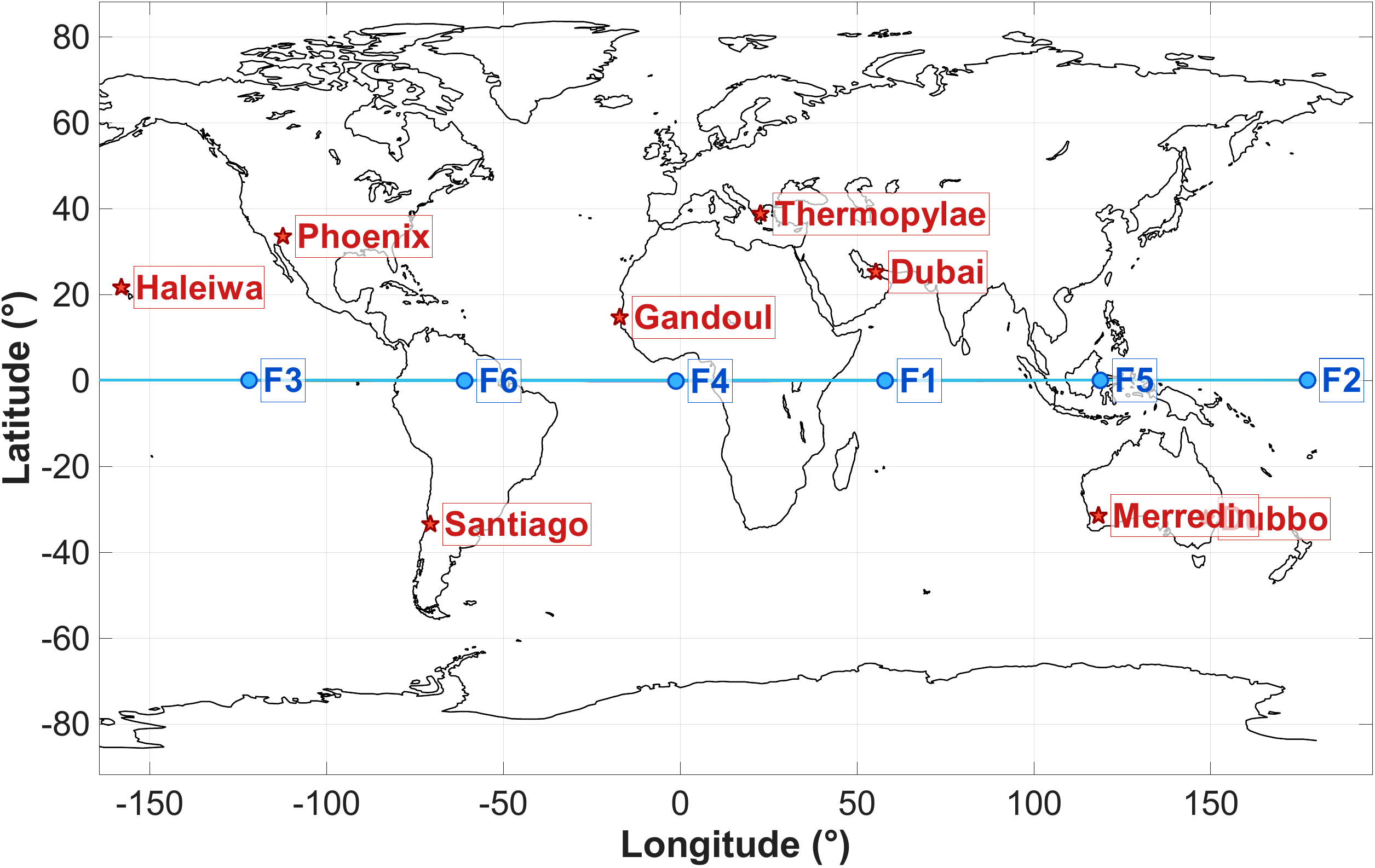}
  \caption{\footnotesize SES O3b mPOWER system architecture showing the equatorial orbit (blue line). Red stars indicate ground stations, and blue dots mark satellite positions at a given time instant.}
  \label{fig:ses-system} \vspace{-6mm}
\end{figure}

To evaluate the performance of the proposed approach for the ACMCFP, simulations were developed using the satellite communications toolbox of MATLAB, which provides tools for modeling and analyzing satellite constellations, ground stations, and communication links based on real orbital data. In this simulation setup, the SES O3b mPOWER MEO\,constellation is represented using two-line element~(TLE) files to propagate in realistic MEO trajectories at around 8000 km altitude. The O3b mPOWER system is operated by SES, which is designed to deliver high-throughput, low-latency broadband connectivity across the globe. Six satellites (F01–F06) were included in the simulation, as they are currently operational, as shown in Fig. \ref{fig:ses-system}. The ground segment consists of eight GS\,located in Australia (Dubbo, Merredin), Greece (Thermopylae), the United States (Phoenix, Hawaii), Chile (Santiago), UAE (Dubai), and Senegal (Gandoul), positioned to maintain near-global visibility with the orbiting satellites. Table \ref{tab:param-config} summarizes the simulation parameters.

\begin{figure*}[t]   % or [htbp]
  \centering
\resizebox{\textwidth}{0.42\textwidth}{%
  \includegraphics{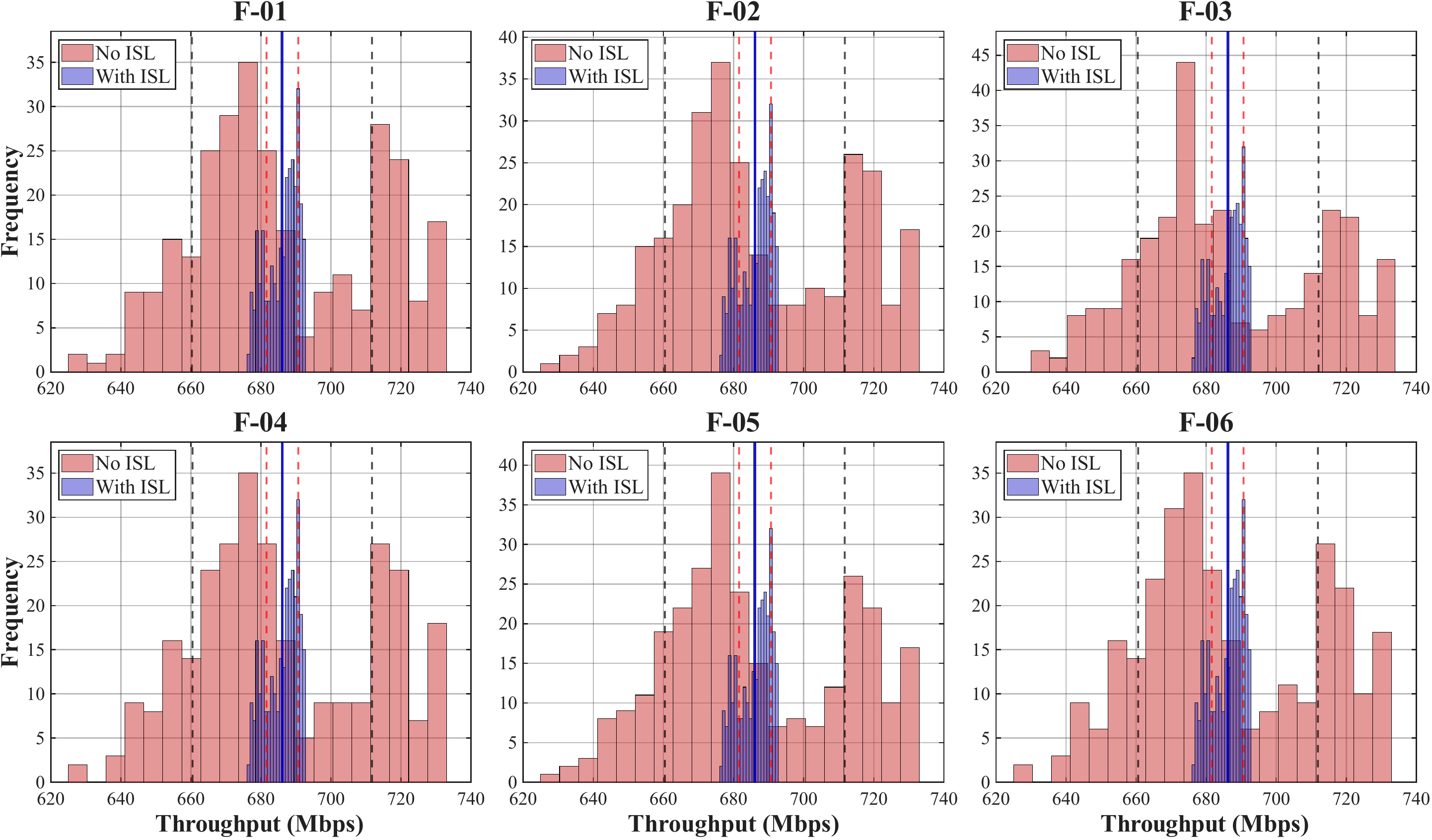}%
}\caption{\footnotesize Throughput distribution histograms for satellites F-01 to F-06 under clear sky conditions. For each satellite, red histograms represent the throughput distribution without inter-satellite links (ISL), while blue histograms correspond to the case with ISL enabled. For each satellite, the solid blue vertical line indicates the mean throughput, whereas the dashed vertical lines denote standard deviation above and below the mean for the respective no-ISL and ISL cases.}
  \label{fig:Histograms}
\end{figure*}

Fig. \ref{fig:Histograms} illustrates the per-satellite throughput distributions for F-01 to F-06 under no rain conditions, comparing cases without ISL and with ISL. In the no-ISL scenario, the histograms exhibit a larger distribution spanning approximately 625 to 735~Mbps, with mean data rates per individual satellite in the order of 686.0 Mbps, and standard deviations around 26.0 Mbps. This widespreadness reflects substantial throughput variability, as satellites experience different FL channel conditions independently. In contrast, with ISL enabled, all distributions collapse into a remarkably narrow peak centered tightly around 686~Mbps with a dramatically reduced standard deviation of approximately 4.6 Mbps. The ISL cooperation thus maintains the ensemble-average throughput while achieving an 82\% reduction in variance, demonstrating that ISL load balancing delivers both steadier per-satellite performance and fairness across the constellation.

\begin{figure*}[t]   % or [htbp]
  \centering
\resizebox{\textwidth}{0.42\textwidth}{%
  \includegraphics{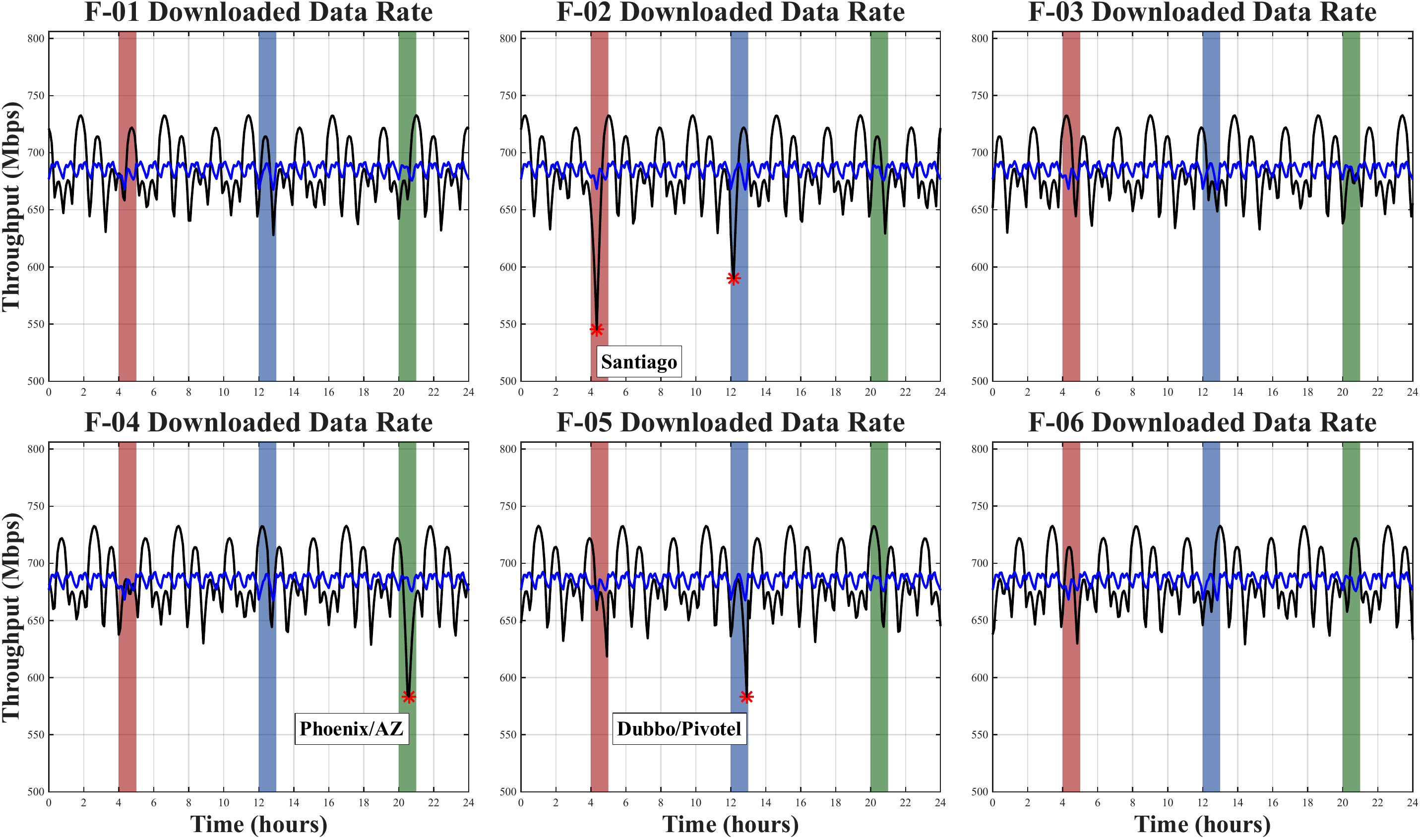}%
}\caption{\footnotesize Downloaded data rate comparison for the six satellites in a $24$-hour window. The light red, light blue, and light green patches indicate heavy rain at Santiago GS, moderate rain at Dubbo/Pivotel GS, and light rain at Phoenix/AZ GS, respectively. The black curves show the no-ISL throughput, with red markers showing the worst-case values under different rain conditions, whereas the blue curves indicate the throughput achieved with ISL.}
  \label{fig:Rainfigure}\vspace{-2mm}
\end{figure*}

Fig. \ref{fig:Rainfigure} shows the per-satellite downlink throughput over a 24-hour window (5-min resolution) for satellites F-01 to F-06. This plot compares scenarios with and without ISL cooperation in the presence of varying rain intensities at different times and ground stations. Without ISLs (black lines), the baseline feeder-link performance exhibits strong temporal fluctuations and frequent drops, particularly during rain events. Heavy rain at Santiago (shown in red shading in F-02, between 4:00 and 5:00 UTC) causes the most severe degradation, reducing throughput to around 545 Mbps. Moderate rain at Dubbo/Pivotel started between 12:00 and 13:00 UTC (blue shading in F-02); however, F-02 was not too much affected, while the FL of F-05 results in more attenuation, dropping the throughput to approximately 583 Mbps. Similarly, the light rain at Phoenix/AZ (green shading in F-04, between 20:00 and 21:00 UTC) produces milder degradation. In contrast, ISL-enabled routing (blue lines) yields a much smoother throughput profile, maintaining higher minimum rates throughout the day at approximately 680–690 Mbps. The ISL capability allows traffic rerouting through unaffected satellites during rain events, effectively mitigating localized feeder-link degradation. 
\begin{table}[t]
\caption{\footnotesize Simulation Parameters for FL and Optical ISL}
\label{tab:param-config}
\centering
\footnotesize
\renewcommand{\arraystretch}{1.1}
\setlength{\tabcolsep}{6pt}
\begin{tabular}{|p{0.40\linewidth}|p{0.18\linewidth}|p{0.12\linewidth}|p{0.10\linewidth}|}
\hline
\textbf{Parameter} & \textbf{Symbol} & \textbf{Units} & \textbf{Value} \\
\hline\hline
\multicolumn{4}{|c|}{\textbf{RF Feeder Link (FL) Parameters}}\\
\hline
Downlink frequency & $f_{\mathrm{DL}}$ & GHz & 20 \\
\hline
Uplink frequency & $f_{\mathrm{UL}}$ & GHz & 29.5 \\
\hline
Downlink bandwidth & $B_{\mathrm{DL}}$ & MHz & 100 \\
\hline
Satellite EIRP & $\mathrm{EIRP}_{\mathrm{sat}}$ & dBW & 49.7 \\
\hline
Satellite $G/T$ (DL) & $(G/T)_{\mathrm{sat,DL}}$ & dB/K & 7 \\
\hline
Satellite $G/T$ (ISL) & $(G/T)_{\mathrm{sat,ISL}}$ & dB/K & 12 \\
\hline
GS antenna diameter & $D_{\mathrm{GS}}$ & m & 4.5 \\
\hline
GS system temperature & $T_{\mathrm{sys}}$ & K & 150 \\
\hline
Minimum elevation & $E_{\min}$ & deg & 5 \\
\hline\hline
\multicolumn{4}{|c|}{\textbf{Optical Inter-satellite Link (ISL) Parameters}}\\
\hline
Laser wavelength & $\lambda$ & nm & 1550 \\
\hline
Transmitter optical efficiency & $\eta_k$ & -- & 0.8 \\
\hline
Receiver optical efficiency & $\eta_\ell$ & -- & 0.8 \\
\hline
Receiver telescope diameter & $D_\ell$ & mm & 80 \\
\hline
Transmitter pointing error & $\theta_k$ & $\mu$rad & 1 \\
\hline
Receiver pointing error & $\theta_\ell$ & $\mu$rad & 1 \\
\hline
Full transmitting divergence angle & $\Theta_k$ & $\mu$rad & 15 \\
\hline
Receiver sensitivity & $P_{\mathrm{req}}$ & dBm & $-35.5$ \\
\hline
\end{tabular} \vspace{-2mm}
\end{table}
\vspace{-4mm}
\section{Conclusion}
This paper derived a relevant max–min fairness optimization problem for MEO constellation with optical ISL. The problem was modelled as an ACMCFP that maximizes the minimum per-satellite download rate and solved it with a per-slot linear program. The formulation offers the possibility to identify the amount of data that should be downloaded to the different GS endpoints (anycast) and shift load over adjacent ISLs, while constraints enforce flow conservation and per-link capacity limits. The linear program, whose solution directly specifies FL and ISL-rate allocations and per-slot routes, was then solved. The evaluations indicate that coordinated ISL use improves the minimum per-satellite performance instantaneous data rate and stabilizes service without additional resources. 
Overall, the proposed ACMCF formulation provides a practical foundation for fairness-driven routing and resource allocation in integrated space–ground networks.
\vspace{-2mm}
\bibliographystyle{IEEEtran}
\bibliography{Final_References.bib}

\end{document}